\documentclass[aps,prd,fleqn,showpacs,showkeys,Nofootinbib,twocolumn,
fleqn,preprintnumbers]{revtex4}
\usepackage{graphicx}
\usepackage{bm}
\usepackage{amsmath}
\usepackage{amssymb}
\usepackage{amsfonts}

\usepackage[normalem]{ulem} 
\usepackage[dvips]{color} 
\renewcommand\sout{\bgroup \color{red} \ULdepth=-.5ex \ULset}

\newcommand{\Slash}[1]{\ooalign{\hfil/\hfil\cr$#1$}} 
\date{\today}

\begin{document}
\preprint{INHA-NTG-05/2014}
\title{Stability of the pion and the pattern of chiral symmetry breaking}

\author{Hyeon-Dong Son}

\email{hdson@inha.edu}

\affiliation{Department of Physics, Inha University, Incheon 402-751,
  Republic of Korea}
\author{Hyun-Chul Kim}

\email{hchkim@inha.ac.kr}

\affiliation{Department of Physics, Inha University, Incheon 402-751,
  Republic of Korea}
\affiliation{Research Center for Nuclear Physics (RCNP), 
Osaka University, Ibaraki, Osaka, 567-0047, Japan}

\begin{abstract}
We investigate the pressure of the pion, which should be equal to zero
to ensure the stability of the pion, within the framework of the
chiral quark model beyond the chiral limit. The pressure of the pion
turns out to vanish nontrivially by the Gell-Mann-Oakes-Renner
relation within the present framework. It implies that the stability
of the pion might be deeply rooted in spontaneous chiral symmetry
breaking. We also discuss physical quantities relevant to the
energy-momentum tensor operator. 
\end{abstract}


\maketitle

\textbf{1.} The energy-momentum tensor form factors (EMTFFs) of
hadrons, also known as the gravitational form factors, provide crucial
information on how quarks and gluons are distributed to form a
hadron\cite{Pagels:1966zza}. On the one hand, it is almost impossible
to get access to the EMTFFs experimentally, because they are probed by
very weak graviton exchange. On the other hand, the EMTFFs are related
to the moments of the generalized parton distributions
(GPDs)\cite{Mueller:1998fv,Ji:1996ek, Radyushkin:1996nd,
Goeke:2001tz,  Diehl:2003ny}. They are identified as the second
moments of the isoscalar GPDs that are defined as the hadronic matrix
elements of a nonlocal flavor-singlet vector operator. These GPDs are
experimentally accessible, for example, via deeply virtual Compton
scattering or exclusive reactions.

The EMTFFs of the pion reveal explicitly its internal structure, in
particular, in regard to spontaneous chiral symmetry breaking
(S$\chi$SB), since the pion, the lightest hadron, is a Goldstone boson
arising from S$\chi$SB. The diagonal space components of the
energy-momentum tensor (EMT) are identified as the pressure of
the pion which should vanish to ensure its stability. This stability
condition of the pion should be satisfied by any approach of
describing the structure of the pion. In this Letter, we want to show
how the pressure of the pion is just expressed in terms of the
Gell-Mann-Oakes-Renner (GOR) relation~\cite{GellMann:1968rz}, based on 
the chiral quark model ($\chi$QM). The model can be constructed in
such a way that chiral symmetry and its spontaneous breaking is
realized by the nonlinear pion field without an elementary scalar
field. The $\chi$QM furnishes a simple but effective framework in
investigating the structure of the pion. It has been shown to be
successful in explaining various properties of the pion. The EMTFFs of
the pion have been already studied within similar frameworks in the
chiral limit. In this limit, the pressure of the pion trivially
vanishes. When one turns on chiral symmetry breaking explicitly,
prominent features emerge: The pion mass leads to a split of the two
EMTFFs at zero momentum transfer and the stability of the pion is
secured by the GOR relation. Both features are all deeply rooted in
the pattern of explicit chiral symmetry breaking.
\vspace{0.1cm}

\textbf{2.} The isoscalar vector GPD of the pion is defined as the
pionic matrix element of the nonlocal vector
current~\cite{Polyakov:1999gs,Diehl:2005rn}
\begin{widetext}
\begin{equation}
  \label{eq:2}
2 \delta^{ab} H_\pi^{I=0}(x,\xi,t) \;=\; \frac12 \int \frac{d\lambda}{2\pi}
e^{ix\lambda (P\cdot n)}   \left\langle \pi^a(p') |
\bar{\psi}(-\lambda n/2) \Slash{n} [-\lambda n/2,\lambda n/2]
\psi(\lambda n/2) |\pi^b(p)\right\rangle,
\end{equation}
\end{widetext}
where $n$ denotes the light-like auxiliary vector. The $P$ is defined
as $P=(p'+p)/2$ and $t$ designates the square of the momentum transfer  
$t=(p'-p)^2$. The gauge connection $[-\lambda n/2,\lambda
n/2]$ can be suppressed in the light-cone gauge. While the first
moment of the isovector GPD yields the electromagnetic form factor of
the pion, the second moment of the isoscalar GPD defined in
Eq.~(\ref{eq:2}) provides the generalized form factors $A_{2,0}$ and
$A_{2,2}$ as
\begin{equation}
  \label{eq:DefGPD1}
  \int dx\, x  H_\pi^{I=0} (x,\,\xi,t) \;=\; A_{2,0}(t) + 4 \xi^2 A_{2,2} (t).
\end{equation}
These two form factors are in fact the same as the EMTFFs,
which are defined as the matrix element of the EMT
\begin{eqnarray}
  \label{eq:matrixelement}
&&	\left\langle \pi^a(p') | T_{\mu\nu}(0) | \pi^b(p)\right\rangle \cr
&=&\frac{\delta^{ab}}{2}\left[ (t g_{\mu\nu} - q_\mu q_\nu)
       \Theta_1(t) + 2 P_\mu P _\nu \Theta_2(t) \right],
\end{eqnarray}
where $T_{\mu\nu}$ denotes the quark part of the QCD
EMT operator defined as
\begin{equation}
  \label{eq:EMT4}
T_{\mu\nu}(x) =\frac12 \bar{\psi}(x)\gamma_{\{\mu} i
\overleftrightarrow{\partial}_\nu {}_{\}} \psi(x).
\end{equation}
The form factors $\Theta_1$ and $\Theta_2$ are called the EMTFFs,
which were already put forward by Pagels~\cite{Pagels:1966zza} decades
ago.  Compared with Eq.~(\ref{eq:DefGPD1}), we have the relations
$\Theta_1=-4A_{2,2}^{I=0}$ and $\Theta_2 = A_{2,0}^{I=0}$.
The low-energy theorem tells that the EMTFFs should satisfy the
following two condition at the zero-momentum transfer:
$\Theta_2(0) \;=\;1$ and $\Theta_1(0)-\Theta_2(0) \;=\;
\mathcal{O}(m_\pi^2)$~\cite{Donoghue:1991qv, Polyakov:1999gs}. The
first condition ensures that the pion mass is correctly reproduced,
which is given by the the fully temporal component of the matrix
element of the EMT
\begin{equation}
\label{eq:pionmass}
\left. \left\langle \pi^a(p) | T_{44}(0) | \pi^b(p)\right\rangle \right|_{t =0}
	\;=\;  - 2m_{\pi}^2 \Theta_{2}(0) \delta^{ab},
\end{equation}
which gives the mass of the pion. On the other hand, the pressure of
the pion is defined as the matrix element for the sum of the EMT
spatial components
\begin{equation}
\label{eq:stability}
\left.\left\langle \pi^a(p) | T_{ii}(0) | \pi^b(p)\right\rangle
\right |_{t = 0} \;=\; \left. \frac{3}{2}t \,\Theta_1(t) \right|_{t=0}.
\end{equation}
We will soon show that Eq.~(\ref{eq:stability}) will be related to the
GOR relation. 

We are now in a position to compute the EMTFFs within the framework of 
the $\chi$QM~\cite{Manohar:1983md, Diakonov:1997sj,
  Diakonov:1985eg}. It is known that the $\mathrm{SU(2)}_L\times
\mathrm{SU(2)}_R$ chiral symmetry is spontaneously broken to the
vector subgroup $\mathrm{SU(2)}_V$ by the quark condensate, which
gives rise to the Goldstone bosons that correspond to the homogeneous
space $\mathrm{SU(2)}_L\times \mathrm{SU(2)}_R/\mathrm{SU(2)}_V$ known
as the Goldstone-boson manifold. The Goldstone boson fields or the
pion fields $\Sigma$ satisfy the following transformation $\Sigma\to L
\Sigma R^\dagger$ with the left-handed (right-handed) transformations 
$L(R)$. Thus, having integrated out the quark fields, the
effective chiral action in Euclidean space can be written as follows
  \begin{equation}
  \label{eq:action}
 	S_{\mathrm{eff}} \;=\; -N_c\mathrm{Tr} \log
 	\left[ i \Slash{\partial} + i M \Sigma P_L + i M
          \Sigma P_R + i \hat{m} \right],
  \end{equation}
where $N_c$ is the number of colors and $P_L(P_R)$ denote the
projection operators defined as $P_L=(1-\gamma_5)/2$ and
$P_R=(1+\gamma_5)/2$. The $\mathrm{Tr}$
stands for the functional trace over all involved spaces. The
pseudo-Goldstone boson field $\bm \pi$ is represented nonlinearly as
  \begin{eqnarray}
  	\label{eq:pseudoGB}
 	\Sigma \;=\; \exp \left(\frac{i {\bm \tau}\cdot{\bm \pi}}{f_\pi}\right)
  \end{eqnarray}
with the Pauli matrices $\tau_i$ and the pion decay constant
$f_\pi$. The $M$ is the diagonal vacuum expectation value of the
$\Sigma$, which is called the dynamical quark mass that arises from
S$\chi$SB. Since we consider explicitly the chiral symmetry
breaking~\cite{GellMann:1968rz, Glashow:1967rx}, the current quark
mass matrix $\hat{m}=\mathrm{diag}(m_{\mathrm{u}},\, 
m_{\mathrm{d}})$ is introduced in the effective chiral
action. Assuming isospin symmetry, we can define the current quark
mass $m$ as $m=m_{\mathrm{u}} = m_{\mathrm{d}}$.  

The quark EMT operator corresponding to Eq.~(\ref{eq:action}) has the
same form as Eq.~(\ref{eq:EMT4}). One can now straightforwardly
compute the matrix elements of the EMT
operator:
\begin{eqnarray}
  \label{eq:mtresult1}
&&  \left\langle \pi^a(p_f) | T_{\mu\nu}(0) |
  \pi^b(p_i)\right\rangle \cr
&=& \delta^{ab}\frac{2 N_c}{f_\pi^2} \int d\tilde{k}\,
 \sum_i \mathcal{F}_i(k,p,q)_{\mu\nu}+ (\mu \leftrightarrow \nu)
 \end{eqnarray}
with
 \begin{eqnarray}
   \label{eq:mtresult2}
 \mathcal{F}_{a \mu\nu}&=&
  -\frac{M \overline{M} k_{d\mu} k_{d\nu}}{D_b D_c}\cr
 \mathcal{F}_{b \mu\nu}&=&
 \frac{2M^2 k_{d\nu}}{D_aD_b D_c}
   \left[- k_{a\mu}\left(k_{bc} + \overline{M}^2\right)
  +k_{b\mu}\left(k_{ac} + \overline{M}^2\right) \right. \cr
&+&\left.  k_{c\mu}\left(k_{ab} +\overline{M}^2\right)\right],
\end{eqnarray}
where the internal quark momenta are
defined as $k_{a\mu} = k_\mu - p_\mu/2 - q_\mu/2$, $k_{b\mu} = k_\mu +
p_\mu/2 - q_\mu/2$, and $k_{c\mu} = k_\mu + p_\mu/2 + q_\mu/2$.
In addition, we have introduced abbreviations
$d\tilde{k}=d^4k/(2\pi)^4$, $\overline{M}=m+M$, $D_i = k_i^2 +
\overline{M}^2$, $k_d = k_b + k_c$, and $k_{ij}=k_i\cdot k_j$ for
convenience. It is interesting to observe that redundant quadratic
divergences in Eq.~(\ref{eq:mtresult1}) are canceled each other except
for the term yielding the quark condensate. In the chiral limit, the
quadratic divergent terms are all canceled out.

Selecting the diagonal spatial parts of the matrix elements at zero
momentum transfer and summing them, we arrive at the following
expression
\begin{widetext}
\begin{eqnarray}
\mathcal{P} &=& \left\langle \pi^a(p) | T_{ii}(0) | \pi^a(p)\right\rangle \cr
&=& \frac{12 N_cm M}{f_\pi^2} \int d\tilde{l} \,
\frac{-l^2}{[l^2 +\overline{M}^2]^2} \,
+  \frac{12N_cM^2}{f_\pi^2} \int d\tilde{l}  \int^1_0 dx\,
 \frac{- p^2 l^2}{[l^2+x(1-x)p^2+\overline{M}^2]^3}
\label{eq:zeromt1}
\end{eqnarray}
\end{widetext}
In the chiral limit, the pressure trivially vanishes because of $m=0$
and $p^2=-m_{\pi}^2=0$. When one switches on the explicit chiral
symmetry breaking, Eq.~(\ref{eq:zeromt1}) looks apparently
persistent. However, the first term is pertinent to the quark
condensate in Euclidean space $-i\langle \psi^\dagger \psi\rangle$,
which is related to that in Minkowski space $\langle
\bar{\psi}\psi\rangle=-i\langle \psi^\dagger \psi\rangle$, defined as  
\begin{equation}
\label{eq:QuarkCondensate}
i\langle \psi^\dagger \psi\rangle =
8 N_c\int d\tilde{l} \frac{
  \overline{M}}{[l^2 + \overline{M}^2]},  
\end{equation}
while the second term is proportional to the pion decay
constant expressed as
\begin{equation}
  \label{eq:PionDecayConstant}
f_\pi^2 = 4 N_c  \int^1_0 dx \int d\tilde{l}\,
  \frac{ M\overline{M} }{[l^2 + \overline{M}^2+x(1-x)p^2]^2}.
\end{equation}
Equations (\ref{eq:QuarkCondensate}) and (\ref{eq:PionDecayConstant})
being used, Eq.~(\ref{eq:zeromt1}) turns out to be related to the GOR
relation 
\begin{equation}
\label{eq:zeromt_pressure}
\mathcal{P} \;=\;
 \frac{3M}{  f_\pi^2\overline{M} }
  \left( m \left\langle \bar{\psi} \psi \right\rangle +m_\pi^2 f_\pi^2
  \right), 
\end{equation}
which vanishes. Thus, the pressure of the pion beyond the chiral limit
is still kept to be zero and consequently the stability of the pion is
ensured. This is a remarkable result, since it gives a hint that the
stability of the pion should be deeply rooted in $S\chi$SB and the
pattern of explicit chiral symmetry breaking. The result of 
Eq.~(\ref{eq:zeromt_pressure}) is unique for the pion.  
\vspace{0.2cm}

\textbf{3.} We now continue to compute and discuss the EMTFFs and the
corresponding transverse charge densities. The EMTFFs are derived
from the matrix element given in Eq. (\ref{eq:mtresult1})
\begin{widetext}
\begin{eqnarray}
\Theta_1(t)&=&
\frac{1}{t} \frac{4 N_c}{f_{\pi}^2}\int d\tilde{k}
\left[-\frac{M \overline{M}}{D_b D_c}
\left(k_d^2 -\frac{(k_d \cdot P)^2}{P^2}\right)
    +\frac{2M^2}{D_a D_b D_c}\left(
    -(k_{bc} + \overline{M}^2) \left(  k_{ad} - \frac{(k_d\cdot
        P)(k_a\cdot P)}{P^2}\right) \right.\right.\cr
    &&\quad\left.\left.
    +(k_{ac} + \overline{M}^2)\left(  k_{bd} -  \frac{(k_d\cdot
        P)(k_b\cdot P)}{P^2}\right)
    + (k_{ab} + \overline{M}^2)\left(  k_{cd} -
 \frac{(k_d\cdot P)(k_c\cdot P)}{P^2}\right)
    \right)\right], \label{eq:emt_ffs1} \\
\Theta_2(t)&=&
\frac{1}{P^2} \frac{2 N_c}{f_{\pi}^2}\int d\tilde{k}
\left[-\frac{M \overline{M}}{D_b D_c}
\left(-k_d^2 +\frac{3(k_d \cdot P)^2}{P^2}\right)
    +\frac{2M^2}{D_a D_b D_c}\left(
    -(k_{bc} + \overline{M}^2) \left(  -k_{ad} +
    \frac{3(k_d\cdot P)(k_a\cdot P)}{P^2}\right) \right.\right.\cr
    &&\quad\left.\left.
    +(k_{ac} + \overline{M}^2)\left(  -k_{bd} +
 \frac{3(k_d\cdot P)(k_b\cdot P)}{P^2}\right)
    +(k_{ab} + \overline{M}^2)\left(  -k_{cd} +
 \frac{3(k_d\cdot P)(k_c\cdot P)}{P^2}\right)
    \right)\right].
\label{eq:emt_ffs2}
\end{eqnarray}
\end{widetext}
Since Eqs.~(\ref{eq:emt_ffs1}) and (\ref{eq:emt_ffs2}) for the EMTFFs
are divergent, we need to introduce a regularization to tame the
divergences. The proper-time regularization is employed, in which the
current quark mass $m$ and the UV cut-off parameter $\Lambda$ are
fixed to be $m=16\,\mathrm{MeV}$ and $\Lambda=652\,\mathrm{MeV}$,
respectively, by using the experimental data $m_{\pi}=139.57$ MeV and
$f_{\pi}=93\,\mathrm{MeV}$. Note that the value of $\Lambda$ is not
much different from that in the chiral limit
($\Lambda=650\,\mathrm{MeV}$). The quark condensate turns out to be
$\langle \bar{\psi}\psi\rangle=-(220)^3 \,\mathrm{MeV}^3$. The
dynamical quark mass, which is only the free
parameter of the present model, is taken to be 350
MeV~\cite{Diakonov:1985eg}.

\begin{figure}[ht]
  \centering
  \includegraphics[scale=0.8]{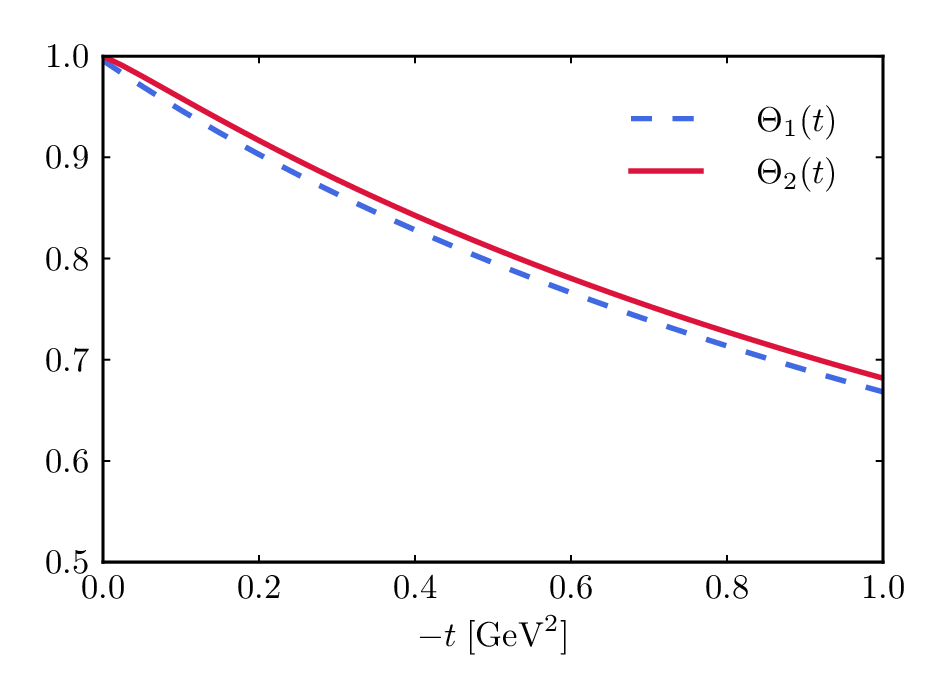}
  \caption{The energy-momentum tensor form factors beyond the chiral
    limit.}
  \label{fig:1}
\end{figure}
Fig.~\ref{fig:1} draws the EMTFFs $\Theta_1(t)$ and $\Theta_2(t)$ as
functions of $t$. As shown in Fig.~\ref{fig:1},
$\Theta_1$ is deviated from $\Theta_2$ with the explicit chiral
symmetry breaking, whereas they are identical in the chiral
limit. Though $t$ increases, the difference between $\Theta_1$
and $\Theta_2$ are kept to be more or less stable.
When the momentum transfer is very small, it was explicitly
shown in chiral perturbation theory ($\chi$PT) that the EMTFFs
are expressed in terms of the gravitational low-energy constants
(LECs) $L_{11}$, $L_{12}$, and $L_{13}$ in the presence of
gravity\cite{Donoghue:1991qv}:
\begin{eqnarray}
\label{eq:lcs}
\Theta_1(t) &=& 1 + \frac2{f_\pi^2} \left[t (4 L_{11}+L_{12})
-8m_{\pi}^2(L_{11} - L_{13}) \right]\cr
\Theta_2(t) &=& 1 -  \frac{2t}{f_\pi^2}  L_{12}.
\end{eqnarray}
Expanding Eqs.(\ref{eq:emt_ffs1}) and (\ref{eq:emt_ffs2}) with respect
to $t$ or the chiral effective action~(\ref{eq:action}) in curved
space by the heat-kernel expansion~\cite{Ball:1988xg,
  Vassilevich:2003xt}, we find the values of the
LECs to be
\begin{eqnarray}
L_{11} &=&  \frac{N_c}{192\pi^2}=1.6\times 10^{-3},\cr
L_{12} &=& -2L_{11}=-3.2\times 10^{-3},\cr
L_{13} &=& -\frac{N_c}{96\pi^2}\frac{M}{B_0}\Gamma
\left(0,\frac{M^2}{\Lambda^2}\right)=0.84\times10^{-3},
\end{eqnarray}
where $\Gamma(0,M^2/\Lambda^2)$ stands for the incomplete Gamma
function with the cutoff mass $\Lambda$ and $B_0=-\langle
\bar{\psi}{\psi}\rangle/f_\pi^2$. The results of the LECs are in good
agreement with those from $\chi$PT given as $L_{11} = 1.4\times
10^{-3}$, $L_{12}=-2.7\times 10^{-3}$, and $L_{13}=0.9\times
10^{-3}$ at the scale of 1 GeV. The identical expressions for $L_{11}$
and $L_{12}$ were already derived in
Refs.~\cite{Andrianov:1998fr,Megias:2004uj, Megias:2005fj,
  Broniowski:2008hx}, whereas the value of $L_{13}$ is distinguished
from those in them.

The transverse charge densities provide essential information on how
the quarks inside a pion are distributed. Once the EMTFFs or the
generalized form factors are given, one can derive the transverse
charge density $\rho_{20}(b)$ as a function of the impact parameter
$b$ in the transverse plane by carrying out the two-dimensional
Fourier transform of them:
\begin{eqnarray}\label{eq:TrChD}
\rho_{20}(b) = \int^\infty_0 \frac{QdQ}{2 \pi} J_0(bQ)\Theta_2(t),
\end{eqnarray}
where $b = \sqrt{b_x^2+b_y^2}$.
\begin{figure}[ht]
  \centering
  \includegraphics[scale=1.0]{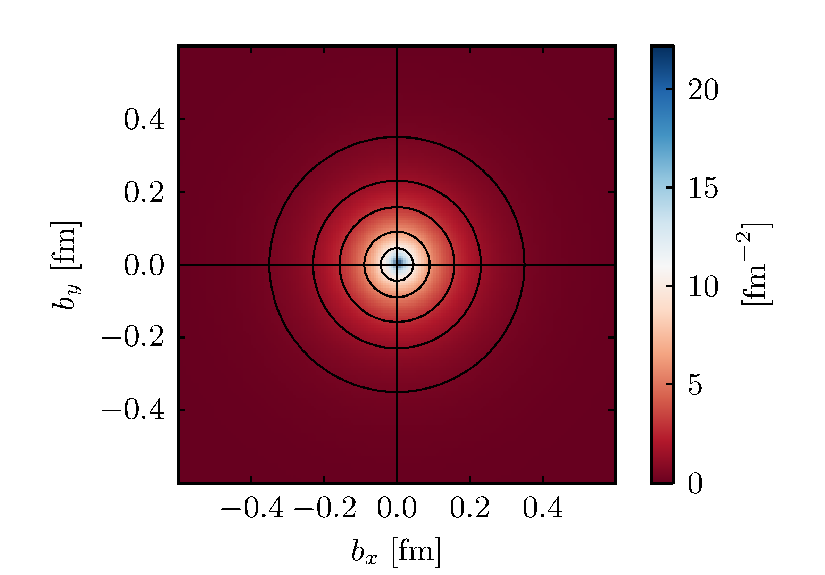}
  \caption{The transverse charge density $\rho_{20}(b)$ as a function
    of $b=\sqrt{b_x^2+b_y^2}$.}
  \label{fig:2}
\end{figure}
\begin{figure}[ht]
  \centering
  \includegraphics[scale=0.8]{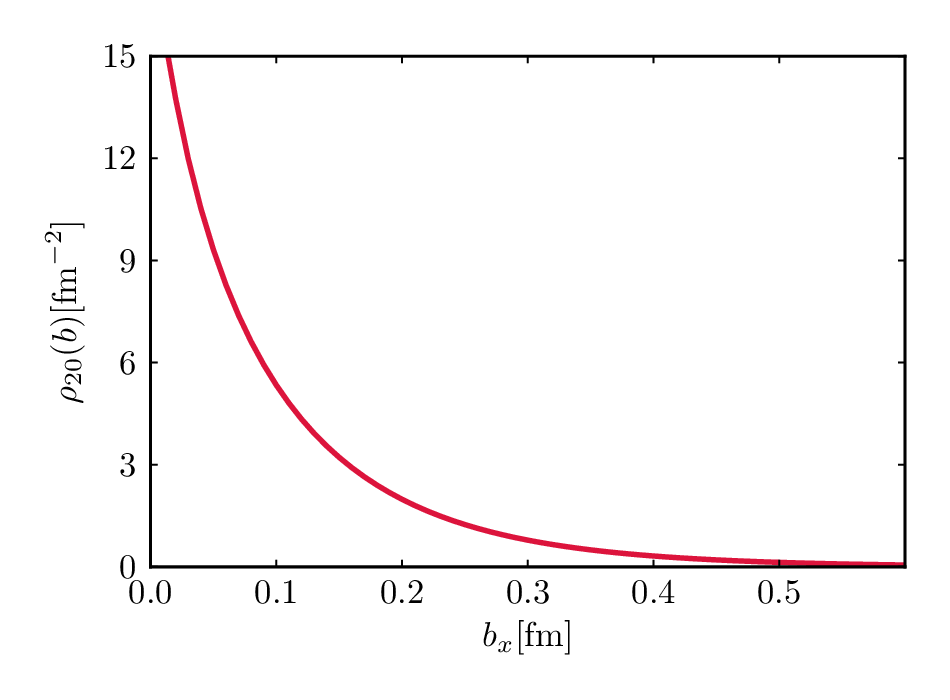}
  \caption{The transverse charge density $\rho_{20}(b)$ as a function
    of $b=\sqrt{b_x^2+b_y^2}$ with $b_y$ fixed to be zero.}
  \label{fig:3}
\end{figure}
In Fig.~\ref{fig:2}, we illustrate the transverse charge density
$\rho_{20}(b)$ of the pion as a function of $b$, which is spherically
symmetric. With $b_y$ fixed to be zero, we draw $\rho_{20}(b)$ in
Fig.~\ref{fig:3}, which shows that $\rho_{20}(b)$ becomes singular as
$b_x$ approaches zero. This singular behavior is already known for the
transverse charge density of the pion in the leading
order~\cite{Miller:2009qu, Vall:2013ina}.

The square of the transverse charge radius of the EMTFFs is defined
as
\begin{eqnarray}\label{eq:rmsr_tr}
	\left\langle b^2\right\rangle_{20} = \int_0^\infty d^2b\, b^2
        \rho_{20}(b).
\end{eqnarray}
We derive the corresponding numerical results as $\sqrt{\langle b^2
  \rangle_{20}} =0.270\,\mathrm{fm}$ and $0.264~\mathrm{fm}$,
respectively, in the chiral limit and with $m_\pi=140$ MeV. These
values are approximately two times smaller than those of the
transverse charge radius of the pion electromagnetic form factor. For
example, we obtain $\sqrt{\langle b^2 \rangle_{10}} =
0.535\,\mathrm{fm}$ with the same pion mass, based on the results of
Ref.~\cite{Nam:2007gf}, which is almost the same as the
phenomenological one  $0.53\,\mathrm{fm}$~\cite{Miller:2010tz}.
\vspace{0.2cm}

\textbf{4.} In this Letter, we investigated the energy-momentum tensor
form factors or the generalized form factors of the pion, based on the
chiral quark model with explicit flavor SU(3) symmetry breaking . We
first showed that the pressure of the pion vanishes due to the
Gell-Mann-Oakes-Renner relation, which is essential to keep the pion
stable. We studied a difference between the two form factors
$\Theta_1(t)$ and $\Theta_2(t)$ at $Q^2=0$, which arises from the
finite pion mass. We presented the results of the low-energy constants
for the energy-momentum tensor form factors of the pion in comparison
with those from chiral perturbation theory. Finally, we computed the
corresponding transverse charge density of the pion electromagnetic
form factor. The present result exhibits a singular behavior as the
impact parameter approaches zero. The present transverse charge
radius turns out to be approximately as twice as smaller than that of
the pion electromagnetic form factor.

\vspace{0.2cm}
The present work is supported by the Grant of Inha University (2014).
H.Ch.K. is grateful to Atsushi Hosaka for valuable discussions and his
hospistality during his stay at RCNP, Osaka University, where part of
the present work has been carried out.

\end{document}